\newcommand{\dir}{./}
\newcommand{\ud}{\mbox{d}}

\newcommand{\etal}{\mbox{\em et al.\ }}

\newcommand{\xu}{\xi_{_0}}
\newcommand{\xh}{\xi_{_{\Gamma}}}

\newcommand{\tres}{\tilde{t}}
\newcommand{\sres}{\tilde{s}}
\newcommand{\Cres}{\tilde{C}}
\newcommand{\Gres}{\tilde{\Gamma}}

\newcommand{\mpsim}{\overline{\Psi}_m}
\newcommand{\mpsis}{\overline{\Psi}_s}
\newcommand{\mphim}{\overline{\Phi}_m}
\newcommand{\mphis}{\overline{\Phi}_s}

\RequirePackage{fix-cm}
\documentclass[twocolumn]{svjour3}          
\smartqed  
\usepackage{graphicx}
\begin{document}

\title{Interplay of curvature-induced micro- and nanodomain structures
in multicomponent lipid bilayers}


\author{Leonie Brodbek         \and
        Friederike Schmid 
}


\institute{L. Brodbek \at
             Institut f\"ur Physik, Johannes Gutenberg-Universit\"at Mainz, DE
           \and
           F. Schmid \at
             Institut f\"ur Physik, Johannes Gutenberg-Universit\"at Mainz, DE
            \email{friederike.schmid@uni-mainz.de}           
}

\date{Received: date / Accepted: date}

\maketitle

\begin{abstract}

We discuss different mechanisms for cur\-vature-induced domain formation in
multicomponent lipid membranes and present a theoretical model that allows us
to study the interplay between the domains.  The model represents the membrane
by two coupled monolayers, which each carry an additional order parameter field
describing the local lipid composition. The spontaneous curvature of each
monolayer is coupled to the local composition; moreover, the lipid compositions
on opposing monolayers are coupled to each other. Using this model, we
calculate the phase behavior of the bilayer in mean-field approximation. The
resulting phase diagrams are surprisingly complex and reveal a variety of
phases and phase transitions, including a decorated microdomain phase where
nanodomains are aligned along the microdomain boundaries. Our results suggest
that external membrane tension can be used to control the lateral
organization of nanodomains (which might be associated with lipid ''rafts'') in
a multicomponent lipid bilayer.

\end{abstract}

\section{Introduction}
\label{intro}

Biomembranes are not homogeneous \cite{VSM03}. They consist of a self-assembled
lipid bilayer which acts as a support for proteins and other biomolecules
\cite{SN72}. In the last decades, there is increasing evidence that
biomembranes are laterally structured \cite{VSM03}, and this structure is
believed to be important for their biological function. For example, the
so-called lipid raft hypothesis states that biomembranes are often filled with
nanodomains, which typically have sizes between 10 to 100 nm, a higher
cholesterol content and higher local order
\cite{ABL97,SI97,BL98,Pike03,Pike06,Leslie11,LS10}. Indirect evidence for the
existence of such rafts is provided, {\em e.g.}, by superresolution images
showing nanoscopic clusters of raft-associated proteins in membranes
\cite{ERM09,MAD11,OMW12}.  However, the main forces driving this internal
organization of membranes have not yet been identified unambiguously.  In
particular, the role of the lipids and the question whether and how they
contribute to the structuring of membranes is still discussed controversially. 

On the one hand, it is argued that the formation of nanoscopic or larger
protein clusters in membranes could be driven by the proteins alone. On the
other hand, nano- and microstructures have also been observed in pure lipid
membranes.  Already one-component phospholipid membranes exhibit a modulated
''ripple phase'' in the transition region between the high temperature fluid
phase and the low-temperature gel phase \cite{KC98,KC02,KTL00,LK02,LS07,SDL14}.
Experimental evidence for nanodomains with properties similar to those
attributed to rafts has been provided by atomic force microscopy \cite{CH13}
and neutron scattering \cite{AMD13,RM13,LPS14,NCM15}. On larger scales,
multicomponent giant vesicles were found to feature ordered patterns of
micron-size domains under certain circumstances \cite{BHW03}. Lateral phase
separation was observed in model multicomponent vesicles \cite{VK03,VK05,VK05b}
and in plasma vesicles extracted from living rat cells \cite{VC08}, and large
up to micron-size critical clusters could be visualized in the vicinity of
critical points \cite{CH13,VC08,VSK07,HCC08,HVK09}.

Several physical mechanism have been proposed that may generate micro- or
nanostructures in pure lipid membranes \cite{KA14}: One, mentioned above, is
critical cluster formation in the vicinity of a demixing phase transition
\cite{CH13,VSK07,HVK09}.  A second suggestion is that line-active molecules in
multicomponent mixtures may reduce the line tension and eventually turn an
originally phase-separated mixture into a microemulsion
\cite{SK04,BPS09,HKA09,YBS10,YS11,HKA12,PS13,PYB14}. In the present paper, we
focus on a third generic class of do\-main-stabilizing mechanism in membranes:
The formation of microemulsions or modulated structures due to
cur\-vature-induced elastic interactions.

\begin{figure}
\centerline{
  \includegraphics[width=0.35\textwidth]{\dir/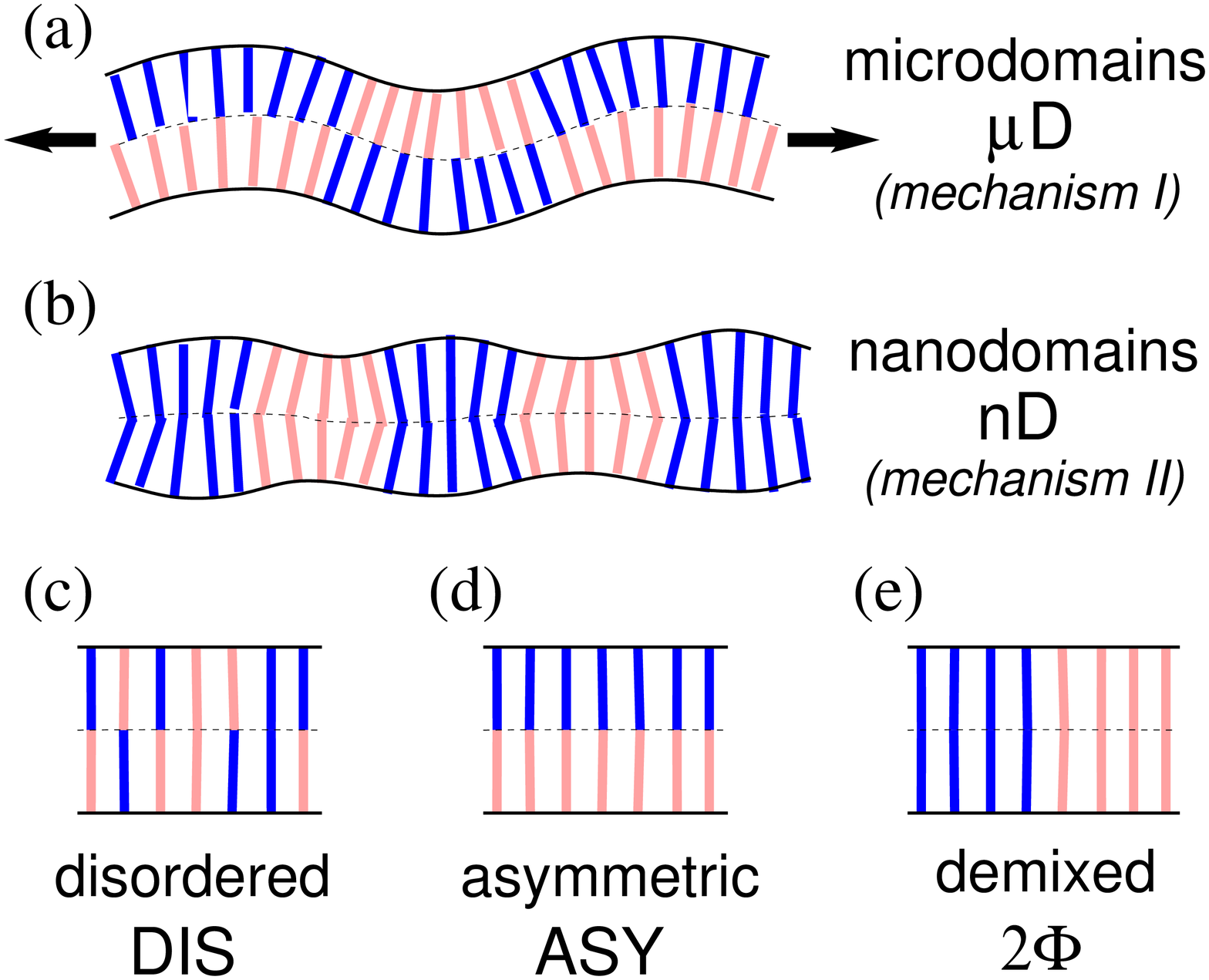}
}
\caption{
Curvature-induced domains in mixed bilayers:
(a) microdomains formed by mechanism I 
\protect\cite{LA87,KGL99,Schick12}
(b) nanodomains formed by mechanism II 
\protect\cite{SDL14,MVS13}.
Panels (c-e) show homogeneous membrane states that
are also considered in this paper:
(c) symmetric disordered state,
(d) asymmetric state
(e) lateral phase separation
}
\label{fig:states}       
\end{figure}

Two variants of such a mechanism have been proposed in the literature.  In the
first (termed I hereafter), it is assumed that the two opposing leaflets tend
to have different lipid composition, and that this imposes a spontaneous
curvature on the {\em bilayer} as a whole \cite{LA87}. A tensionless membrane
responds by bending around \cite{SPA90}.  If tension is applied, it is forced
to be planar, and this creates elastic stress.  The membrane reacts by forming
staggered domains with curvatures of opposite sign. This effect is illustrated
in Fig.\ \ref{fig:states}a). It was first proposed by Andelman and
coworkers and further investigated by a number of authors
\cite{LA87,SPA90,KGL99,HM94,Schick12,SS13,SMS14,SMV14}. Schick and
coworkers argued that it might account for raft formation in membranes
\cite{Schick12,SS13,SMS14}.  The characteristic length scales of the domains
are in the range of 100 $nm$ to micrometers and diverge for tensionless
membranes.

The second mechanism (termed II) was recently proposed in our group
\cite{SDL14,MVS13}.  It assumes that the lipids tend to demix laterally, but
the lipid composition on opposing leaflets are preferably in registry.
Different lipid compositions are associated with different spontaneous
curvature parameters in the {\em monolayer}. Since the membrane as a whole
remains planar, this also creates elastic stress, which is relieved by keeping
domain sizes finite.  The effect is illustrated in Fig.\ \ref{fig:states}b).  
Characteristic length scales according to this mechanism are in the range
of 10 nm and depend on the membrane material. We have argued that the same
effect could also account for the formation of modulated ripple states, which
also have characteristic wave lengths in the same order of magnitude, hence
ripples and nanodomains in lipid bilayers could be closely related phenomena. 

Both mechanisms have in common that they are driven by a coupling between lipid
composition and spontaneous curvature. The main difference lies in the nature
of the local correlation between the lipid compositions on the two leaflets.
Mechanism I assumes that domains are anticorrelated, which favors microdomains.
Mechanism II assumes that domains are in registry, which favors nanodomains.
Experimentally, it has been reported that domains on opposing leaflets tend to
be in registry \cite{CK08}.  However, the correlation is never perfect, and we
shall see below that microdomains may be energetically favorable under certain
circumstances even in systems where the coupling between lipid compositions on
opposing leaflets is weakly positive.

The purpose of the present paper is to provide a unified picture of
curvature-induced phenomena in mixed lipid bilayers. We propose an elastic
theory which reproduces both the mechanisms I and II of domain formation.
Using mean-field approximation, we can assess the influence of membrane
tension, curvature coupling, and composition coupling across leaflets on the
bilayer structure, and calculate a representative set of phase diagrams which
demonstrate the complex interplay between microdomain and nanodomain formation
in these systems. 

Our paper is organized as follows. In the next section, we introduce the
framework of our theoretical approach and derive a Ginzburg-Landau model which
describes both nanodomain and microdomain formation.  In Section
\ref{sec:results} we present and analyze the possible phase diagrams of this
system, which we have calculated analytically within mean-field approximation
and a variational single-mode Ansatz. The results are summarized and discussed
in Section \ref{sec:discussion}. 

\section{Theoretical framework}
\label{sec:model}

Our theory is based on an elastic model for membranes that describes the lipid
bilayer by a system of two coupled elastic monolayers
\cite{DPS93,DBP94,ABD96,BB06}. Each monolayer is represented by a
two-dimensional manifold that describes the position of the monolayer-water
interface in a coarse-grained sense. We assume that the membrane is planar on
average and aligned in the $(x,y)$ plane, and that bubbles and overhangs can be
neglected.  Each monolayer surface can then be parametrized by a function
$z_i(x,y)$ (where $i=1,2$ and $z_1 < z_2$). We consider a symmetric situation
where both monolayers have the same elastic properties, except that their
spontaneous curvatures $c_i$ may vary locally and differ from each other. We
adopt a sign convention according to which $c_i$ is positive if the outer
surface of the monolayer has a tendency to bend inwards, towards the membrane
interior.  Furthermore, we use a Gaussian approximation, {\em i.e.}, we expand
the free energy up to second order of $z_i$ about a fully planar reference
state. The total elastic energy of the coupled monolayer system is then given
by \cite{ABD96,BB06,WBS09,NWN10}

\begin{eqnarray}
\label{eq:f_el}
{\cal F}_{\mbox{\tiny el}} &=&
\int \ud^2 r \: \Big\{
\frac{k_A}{8 t_0^2} (z_1 - z_2)^2
+ \frac{k_c}{4} \big[(\Delta z_1)^2 + (\Delta z_2)^2 \big]
\\ \nonumber
&& \hspace*{2mm}
+\,  k_c \frac{\zeta}{2 t_0} (z_1 - z_2) \big[ \Delta z_1 - \Delta z_2 \big]
\\ \nonumber
&& \hspace*{2mm}
+ \, k_c \big[ c_1 \Delta z_1 - c_2 \Delta z_2 \big]
+ \frac{\Gamma}{2} \big[\frac{1}{2} \nabla (z_1 + z_2)\big]^2
\Big\}
\end{eqnarray}

Here we have not included the contribution of the Gaussian curvature, which is
a constant for planar membranes with fixed topology \cite{Safran94}. The
parameter $t_0$ is the mean monolayer thickness, $k_A$ and $k_c$ are the
(bilayer) compression and bending modulus, respectively, $\zeta$ is a
curvature-related parameter \cite{ABD96,BB06,WBS09}, and $\Gamma$ the 
tension of the membrane \cite{NWN10}. 

We must briefly comment on the interpretation of the parameter $\Gamma$. It has
been noted \cite{FP03} and confirmed by simulations \cite{FB08,Schmid11,SN15}
that the externally applied tension (the ''frame tension'') and the tension
experienced by the lipids inside the membrane (the ''intrinsic'' tension)
differ slightly from each other in fluctuating membranes. However, fundamental
symmetry considerations \cite{CLN94,FP04} suggest that the ''tension''
parameter governing the amplitude of membrane undulations coincides with the
frame tension in the thermodynamic limit, except for a small multiplicative
correction factor that accounts for the difference between the actual area and
the area projected on the $(x,y)$-plane \cite{Diamant11}. This implies that in
a mean-field theory based on a quadratic approximation such as (\ref{eq:f_el}),
the parameter $\Gamma$ is most appropriately interpreted as a frame tension.
Several simulations have confirmed that frame tension and fluctuation tension
are equal \cite{Schmid11,FP04,WF05,NG06,Farago11}, but deviations have also
been reported, especially for low tensions
\cite{FB08,Imparato06,Stecki08,TCB13}. If the membrane tension is close to
zero, the existence of the thermodynamic limit for which the theoretical
arguments \cite{CLN94,FP04} would apply becomes questionable. One consequence
is that the amplitude of undulations may depend on the statistical ensemble
under consideration \cite{Schmid11}. However, these effects are small and we
can neglect them in the context of this work. Hence we identify $\Gamma$ with a
frame tension, which can be applied and controlled externally.

We will also neglect the fact that the material parameters of the membrane may
change under tension \cite{NWN10,AKZ07,WPW13}. Our elastic model does not
explicitly account for the effect of lipid orientation and possible lipid tilt
\cite{Fournier98,Fournier99,BKM03,FIM06,WPW11}. A detailed consideration of
such factors might help to establish a molecular basis for the relation between
the elastic parameters and the lipid structure \cite{KAC05,KHR13}. Here we wish
to keep our model as simple as possible. We emphasize that Eq.\ (\ref{eq:f_el})
contains {\em all} terms up to second order of $z_i$ and the spatial
derivatives that are allowed by symmetry in a system of two coupled identical
monolayers described by two surfaces $z_i$ \cite{Schmid13} (apart from the
contribution of the Gaussian curvature). We have used the elastic model, Eq.\
(\ref{eq:f_el}), to fit deformation profiles in the vicinity of inclusions,
with good results down to molecular length scales \cite{WBS09,NNW12}. 

To describe lateral phase separation and domain formation within the
monolayers, we supply each monolayer with an additional order parameter field
$\varphi_i(x,y)$. Here $\varphi$ is a collective variable designed to
characterize the local lipid composition in a multicomponent system -- not
necessarily a binary system -- with a propensity to locally phase separate. The
associated free energy is described by a Ginzburg-Landau functional

\begin{equation}
\label{eq:f_gl}
{\cal F}_{\mbox{\tiny GL}}
= \int \ud^2 r \: \Big\{ \sum_{i=1}^2 \big[ \frac{g}{2} (\nabla \varphi_i)^2
  + \frac{t}{4} \varphi_i^2 + \frac{v}{8} \varphi_i^4 \big]
  - \frac{s}{2} \varphi_1 \varphi_2
\Big\}
\end{equation}
with $v > 0$. The last term couples the compositions on the two leaflets and is
constructed such that it favors equal composition for $s> 0$ (positive
coupling) and different composition for $s<0$ (negative coupling). In the
absence of any coupling, lateral phase separation occurs for $t < 0$ and the
monolayers remain homogeneous ($\varphi_i \equiv 0$) for $t > 0$. Hence the 
parameter $t$ is temperature-like and describes the distance from the
critical demixing transition in parameter space.

Thus far, we have combined standard theories for fluid elastic sheets and phase
separating order parameter fields. The key additional feature that we must
introduce to describe curvature-induced domain formation is a coupling between
the order parameters $\varphi_i (x,y)$ and the local spontaneous curvature of
the monolayer $c_i(x,y)$. Kollmitzer \etal \cite{KHR13} recently
discussed how the monolayer curvature might depend on the membrane composition
for raft-forming lipids.  Specifically, we assume that the relation between
$c_i$ and $\varphi_i$ is roughly linear,

\begin{equation}
\label{eq:f_coupling}
c_i(x,y) = c_0 + \hat{c} \: \varphi_i(x,y),
\end{equation}
where $c_0$ is the spontaneous curvature in the mixed homogeneous system.  

Eqs.\ (\ref{eq:f_el}) - (\ref{eq:f_coupling}) define our theoretical
model.  To proceed, we express the degrees of freedom in terms of a set of new
fields that make the symmetries in the system more transparent: The position of
the membrane mid-plane, $h(x,y) = \frac{1}{2}(z_1 + z_2)$, the variations of
the local mean monolayer thickness, $u(x,y) = \frac{1}{2} (z_1-z_2) -  t_0$,
the mean local order parameter $\Phi = \frac{1}{2} (\varphi_1 + \varphi_2)$,
and the local order parameter difference between monolayer leaflets $\Psi =
\frac{1}{2} (\varphi_1 - \varphi_2)$.  Rewritten as a functional of these new
fields, the total free energy reads

\begin{eqnarray}
\label{eq:f_tot1}
{\cal F} &=& \int \ud^2 r \: \Big\{
\frac{k_A}{2 t_0^2} u^2 + \frac{k_c}{2} (\Delta u)^2
+ 2 k_c \frac{\zeta}{t_0} u \Delta u
\\ \nonumber
&& \hspace*{2mm}
+ \, \frac{\Gamma}{2} (\nabla h)^2 + \frac{k_c}{2} (\Delta h)^2
\\ \nonumber
&& \hspace*{2mm}
+ \, \frac{g}{2} (\nabla \Phi)^2 + \frac{1}{2} (t-s) \Phi^2 
+ \frac{v}{4} \Phi^4
\\ \nonumber
&& \hspace*{2mm}
+ \, \frac{g}{2} (\nabla \Psi)^2 + \frac{1}{2} (t+s) \Psi^2 
+ \frac{v}{4} \Psi^4
\\ \nonumber
&& \hspace*{2mm}
+ \, 2 k_c \hat{c} \: [\Phi \Delta u + \Psi \Delta h]
+ \frac{3}{2} v \: \Phi^2 \Psi^2
\Big\}.
\end{eqnarray}.

In this representation, it becomes clear that the two sets of fields ($u,\Phi$)
and ($h,\Psi$) are largely independent of each other. The only coupling between
them is introduced by the last term in Eq.\ (\ref{eq:f_tot1}), a nonlinear
fourth order term. The domain forming mechanism I is associated with
modulations in the fields ($h,\Psi$) and mechanism II is associated with
modulations in ($u,\Phi$) (see Fig.\ \ref{fig:states}).

Since the free energy (\ref{eq:f_tot1}) is quadratic in $h$ and $u$, these
degrees of freedom can be integrated out. Alternatively, we may adopt a
mean-field approximation from the outset and minimize the free energy, Eq.\
(\ref{eq:f_tot1}), with respect to $h$ and $u$. Apart from uninteresting
constants, the result is the same and best written in a mixed Fourier- and 
real space representation (omitting constants):

\begin{eqnarray}
\label{eq:f_tot2}
{\cal F} &=& \frac{(2 \pi)^2}{A}
\sum_{\bf q} \Big\{ 
\frac{1}{2} K_{\Phi}\big((\xu q)^2\big) |\Phi_{\bf q}|^2
\\ \nonumber
&& \hspace*{15mm}
+ \frac{1}{2} K_{\Psi}\big((\xh q)^2\big) |\Psi_{\bf q}|^2
\Big \}
\\ \nonumber
&&
+ \frac{v}{4} \, \int \ud^2 r \:
\Big\{
\big(\Phi^4 + \Psi^4\big)
+ 6 \: \Phi^2 \Psi^2
\Big\}
\end{eqnarray}
with
\begin{eqnarray}
\label{eq:k_phi}
K_{\Phi}(x) &=& t-s + \frac{g}{\xu^2} x
 - 4 k_c \hat{c}^2 \frac{x^2}{(1-x)^2 + \Lambda x}
\\
\label{eq:k_psi}
K_{\Psi}(x) &=& t+s + \frac{g}{\xh^2} x
 - 4 k_c \hat{c}^2 \frac{x}{1+x}.
\end{eqnarray}

Here we have introduced the characteristic length scales $\xu = (t_0^2
k_c/k_A)^{1/4}$ and $\xh = (k_c/\Gamma)^{1/2}$ and the dimensionless parameter
$\Lambda = 2 - 4 \zeta \sqrt{k_c/k_A}$. The length scale $\xu$ and the
parameter $\Lambda$ are intrinsic properties of the membrane.  Experimentally,
one finds that the bending rigidity, $k_c$, is roughly proportional to $k_A
t_0^2$, hence $\xu$ should be of the order of the membrane thickness, $2 t_0$.
The parameter $\Lambda$ characterizes the bilayer thickness profiles in the
vicinity of a local perturbation \cite{ABD96,WBS09}: For $\Lambda \ge 4$, they
decay towards the equilibrium thickness in a purely exponential manner; for
$\Lambda < 4$, an oscillatory component emerges, and the membrane becomes
unstable towards deformations at $\Lambda < 0$.  Inserting numbers from
experiments \cite{Marsh06}, all-atom simulations \cite{BB06,LE00,MRY07}, or
coarse-grained simulations \cite{WBS09} for the fluid phase of DPPC bilayers
(dipalmitoyl phosphatidylcholine, of one of the most common lipids in natural
biomembranes), one consistently obtains values around $\xu \sim (0.9-1.4)$ nm,
and $\Lambda \sim 0.7$ \cite{MVS13}. With typical experimental values for
the tension \cite{GMS12,PAF13} in the range of $\Gamma \sim 10^{-5}$N/m,
one finds that $\xh$ is of the order $\sim 100$ nm. It diverges for
tensionless membranes.

\begin{itemize}
\item The parameter $\Lambda$, which characterizes the membrane material.
  It must be positive, otherwise the membrane is not stable
\item The rescaled curvature coupling $\Cres = 2 \hat{c} \: \xu \sqrt{k_c/g}$
\item The rescaled composition coupling $\sres = s \: \xu^2/g$
\item The rescaled temperature parameter $\tres = t \: \xu^2/g$
\item The rescaled tension $\Gres = \Gamma \: t_0 /\sqrt{k_c \: k_A}$
\end{itemize}
In this notation, the characteristic length scale $\xh$ is simply given 
by $\xh = \xu/\sqrt{\Gres}$. 

Our task is to minimize the free energy (\ref{eq:f_tot2}) with
respect to the fields $\Phi$ and $\Psi$, and to calculate the resulting
phase diagrams. This is done in the next section.

\section{Phase behavior of mixed bilayers}
\label{sec:results}

\subsection{Stability analysis}
\label{sec:stability}

At high ''temperature'' $\tres$, the system is disordered,
$\Phi = \Psi \equiv 0$. The disordered state is unstable towards ordering
or phase separation if either $K_{\Phi}(x)$ or $K_{\Psi}(x)$ become
negative for at least one $x>0$. Hence the instability limit is determined
by the maximum value of $t$ at which the minimum or $K_{\Phi}$ or $K_{\Psi}$ 
becomes zero. Four types of instabilities are possible:

\begin{description}

\item[(i)] Instability of $K_{\Phi}(x)$ at $x=0$. This corresponds to an
instability with respect to macroscopic demixing (state $2 \Phi$ in Fig.\
\ref{fig:states}e)

\item[(ii)] Instability of $K_{\Phi}(x)$ at some $x>0$. This corresponds to an
instability with respect to a modulated state characterized by a periodic array
of nanodomains (state nD in Fig.\ \ref{fig:states}b)

\item[(iii)] Instability of $K_{\Psi}(x)$ at $x=0$. This corresponds to an
instability with respect to the formation of a globally asymmetric membrane
(state ASY in Fig.\ \ref{fig:states}d) 

\item[(iv)] Instability of $K_{\Psi}(x)$ at some $x>0$. This corresponds to an
instability with respect to a modulated phase with a periodic array of
microdomains (state $\mu$D in Fig.\ \ref{fig:states}a)

\end{description}

The instabilities of $K_{\Psi}$ ((iii) and (iv)) have been discussed by
Shlomovitz and Schick \cite{SS13}. At low curvature
coupling $\Cres$, the instability (iii) dominates and the membrane becomes
asymmetric at $\tres = -\sres$. At high $\Cres$, the instability (iv) takes
over, and modulated microdomains with a wavelength $\lambda_{\Psi} = 2 \pi \xu
\Gres^{-1/4} (\Cres
- \sqrt{\Gres})^{-1/2}$ emerge for 

\begin{equation}
\label{eq:t_psi}
\tres < \tres_{\Psi} := -\sres +(\Cres - \sqrt{\Gres})^2.
\end{equation}
The two regimes are separated by a Lifshitz critical point at $\Cres =
\sqrt{\Gres}$, at which point the wavelength $\lambda_{\Psi}$ of the
modulations diverges.

The picture that one obtains after analyzing the instabilities of $K_{\Phi}$
((i) and (ii)) is similar, but the resulting scenario differs from that
described above in one important aspect. At low curvature coupling $\Cres$, the
instability (i) dominates and the membrane phase separates at $\tres=\sres$.
At high $\Cres$, the instability (ii) dominates, and a modulated nanodomain
phase emerges. However, the wavelength of this phase remains finite.  The two
regimes are connected by a bicritical point at $\Cres = \sqrt{\Lambda}$
connecting a line of Ising-type transitions (regime (i)) and Brazovskii-type
transitions \cite{Brazovskii75} (regime (ii)), and the wavelength of the
modulated phase at this point is $\lambda_{\Phi} = 2 \pi \xu$. The transition
point $\tres_{\Phi}$ between the modulated and the disordered phase at $\Cres >
\sqrt{\Lambda}$ is defined through a set of implicit equations

\begin{eqnarray}
\epsilon &:=& \frac{\Cres^2}{\Lambda} - 1 =
\frac{\delta^4 + 2 \delta (1+\delta)^2 \Lambda}
 {\Lambda(1+\delta) (\Lambda (1+\delta) - 2 \delta)}
\\
\tres_{\Phi} &=&\sres + 
\frac{\delta (1+\delta) (2+\delta)}{\Lambda (1+\delta) - 2 \delta},
\end{eqnarray}
where $\delta$ is related to the characteristic wavelength $\lambda_{\Phi}$ at
the transition {\em via} $\lambda_{\Phi} = 2 \pi \xu/\sqrt{1+\delta}$ and
vanishes at the bicritical point. Close to the bicritical point, we can expand
both equations in powers of $\delta$, which leads to the expansion

\begin{equation}
\tres_{\Phi} - \sres =
\epsilon \big[
 1 + \frac{\Lambda}{4} (\epsilon - \epsilon^2)
 + \frac{\Lambda}{16} (\Lambda + 4) \epsilon^3
 + {\cal O}(\epsilon^4)
\big].
\end{equation}
Far from the bicritical point, for large $\epsilon$, one obtains the asymptotic
behavior

\begin{equation}
\tres_{\Phi} - \sres \to
\epsilon \: \left\{
\begin{array}{ll}
1+\Lambda/(4-\Lambda) \; & \mbox{for $\Lambda < 2$} \\
\Lambda \; & \mbox{for $\Lambda > 2$}.
\end{array} \right. 
\end{equation}
Hence $\tres_{\Phi} - \sres$ is roughly proportional to $\epsilon$
in the whole range of $\Cres$. Below, we will use the 
approximate expression derived from the leading order of the
series expansion,

\begin{equation}
\label{eq:t_phi}
\tres_{\Phi} \approx \sres + \epsilon = 
\sres + \frac{1}{\Lambda}(\Cres^2 - \Lambda).
\end{equation}
In the total system of coupled order parameters, both the instabilities in
$\Phi$ and $\Psi$ can destroy the homogeneous membrane, and the membrane
enters the phase associated with the instability at highest ''temperature''
$\tres$. Within the ordered or metastable phase, the nonlinear terms in the
free energy expression, Eq. (\ref{eq:f_tot2}), become important, and we must
include them to calculate the full phase diagram. This is done in the next
section.

\subsection{Phase diagrams}
\label{sec:phases}

In the present work, we are not interested in the exact numerical solution 
of the free energy model, (\ref{eq:f_tot1}), but rather in a qualitative
picture of the interplay of the different ordering mechanisms in the membrane.
Therefore, we calculate the mean-field phase diagram within a single mode
approximation which provides us with analytical expressions for the phase
boundaries. Specifically, we make the following Ansatz for the shape of
the order parameter field:

\begin{equation}
\Phi(\vec{r}) = \mphim \cos(k_{\Phi} z) + \mphis, \quad
\Psi(\vec{r}) = \mpsim \cos(k_{\Psi} z) + \mpsis,
\end{equation}
where $k_{\Phi} = 2 \pi/\lambda_{\Phi}$ and $k_{\Psi} = 2 \pi/\lambda_{\Psi}$
are the most unstable wavevectors for given $\Cres$ and $\Gres$ calculated in
Sec.\ \ref{sec:stability}. Inserting this Ansatz in the functional
(\ref{eq:f_tot2}), we obtain the free energy per area

\begin{eqnarray}
\label{eq:f_totsm}
F/A &=& \frac{v}{4} \Big\{
\mphim^2 (\tres-\tres_{\Phi})
+ 2 \mphis^2 (\tres - \sres)
\\ \nonumber
&& \quad
+ \mpsim^2 (\tres - \tres_{\Psi})
+ 2 \mpsis^2 (\tres + \sres)
\\ \nonumber
&& \quad
+ \, \frac{3}{8} (\mphim^4 + \mpsim^4)
 + \mphis^4 + \mpsis^4
 + 3 (\mphim^2 \mphis^2 + \mpsim^2 \mpsis^2)
\\ \nonumber
&& \quad
 + \, \frac{3}{2} \mphim^2 \mpsim^2
 + 3 (\mphim^2 \mpsis^2 + \mphis^2 \mpsim^2)
 + 6 \mphis^2 \mpsis^2)
\Big\},
\end{eqnarray}

which we can now minimize with respect to the amplitudes $\mphim, \mphis,
\mpsim, \mpsis$ in a straightforward manner.  We find that states with minimal
free energy can only sustain one type of order, {\em i.e.}, all amplitudes are
zero except, possibly, one.  Transitions between different states are first
order.

\begin{figure*}
\centerline{
  \includegraphics[width=0.9\textwidth]{\dir/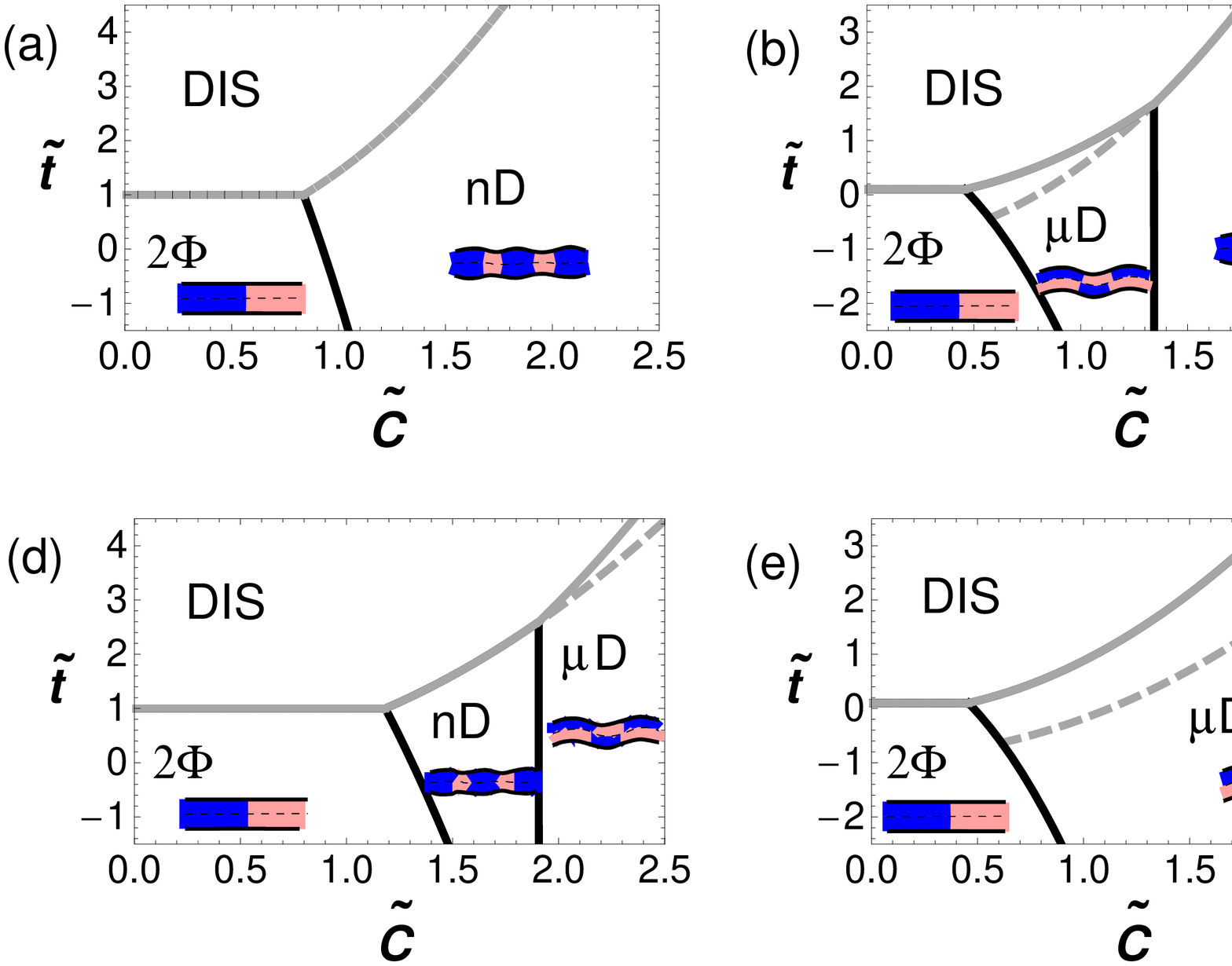}
}
\caption{
Typical phase diagrams in the plane of ''temperature'' $\tres$ and
curvature coupling $\Cres$ for strong, weak, and negative monolayer 
composition coupling ($\sres=1,0.1,-0.1$) and two choices of the membrane 
parameter $\Lambda$, one with $\Lambda < 1$ and one with $\Lambda > 1$:
(a) $\sres=1, \Lambda = 0.7$;
(b) $\sres=0.1, \Lambda = 0.7$;
(c) $\sres=-0.1, \Lambda = 0.7$;
(d) $\sres=1, \Lambda = 1.4$;
(e) $\sres=0.1, \Lambda = 1.4$;
(f) $\sres=-0.1, \Lambda = 1.4$.
The dimensionless tension is $\Gres=10^{-4}$ in all cases, corresponding to a
characteristic length scale for microdomains of $\xh=100 \xu$.  The membrane
structures in the different phases correspond to those shown in Fig.\
\ref{fig:states}.  Solid gray lines correspond to continuous transitions, black
lines to first order transitions, and dashed grey lines transition regions
between a a pure $\mu$D phase (at high $\tres$) and a decorated $\mu$D phase
which also contains nanodomains at the microdomain boundaries (at lower
$\tres$).  The red arrow indicates the position of a Lifshitz point.
}
\label{fig:phdiag_c}       
\end{figure*}

These predictions are consistent with the known behavior near Lifshitz critical
points and seem reasonable in many other aspects as well. However, they clearly
do not capture the interplay of microdomains and nanodomains in the limit
$\xh/\xu \to 0$ or $\Gres \ll 1$  when the characteristic lengths of
microdomains and nanodomains are very different. If the order parameter
$\Psi(\vec{r})$ varies very slowly compared to $\Phi(\vec{r})$, it acts almost
like a constant field on $\Phi$, and it is clear that nanodomains might emerge
in regions with $\Psi \approx 0$ even if the global amplitude of $\Psi$
modulations does not vanish. Hence an adiabatic approximation seems more
appropriate for this case,  where $\mphim$ is allowed to vary in space and to
depend on the local value of $\Psi^2$. 

Let us first assume that we can impose a given constant asymmetry $\Psi$ on a
mixed membrane with $\Phi \equiv 0$.  As long as $\Psi^2 <
\frac{1}{3}(\tres_{\Phi} - \tres)$, the free energy (\ref{eq:f_totsm}) can then
be lowered by introducing a periodic modulation with squared amplitude
$\mphim^2 = - \frac{4}{3} \: (\tres - \tres_{\Phi} + 3 \Psi^2)$. The
associated free energy gain per area is $F_m/A = - \frac{3v}{32} \mphim^4$.  If
we now consider a modulated microdomain phase $\mu$D with very slowly varying
order parameter $\Psi(\vec{r})$, we can lower the free energy by allowing for
the formation of nanodomains at the interfaces between microdomains, and the
associated adiabatic free energy can be approximated as

\begin{equation}
\label{eq:f_ad}
F_{\mbox{\tiny ad}}/A = - \frac{v}{6} \int \ud^2 r \: 
\Big[ \min \big(\tres - \tres_{\Phi} + 3 \Psi(\vec{r})^2, 0 \big) \Big]^2.
\end{equation}
However, the membrane is not entirely filled with nanodomains in this
state, the nanodomains only build up in narrow regions close
to the microdomain boundaries. The transition between the decorated $\mu$D and
the nD phase remains discontinuous.

\begin{figure*}
\centerline{
  \includegraphics[width=0.9\textwidth]{\dir/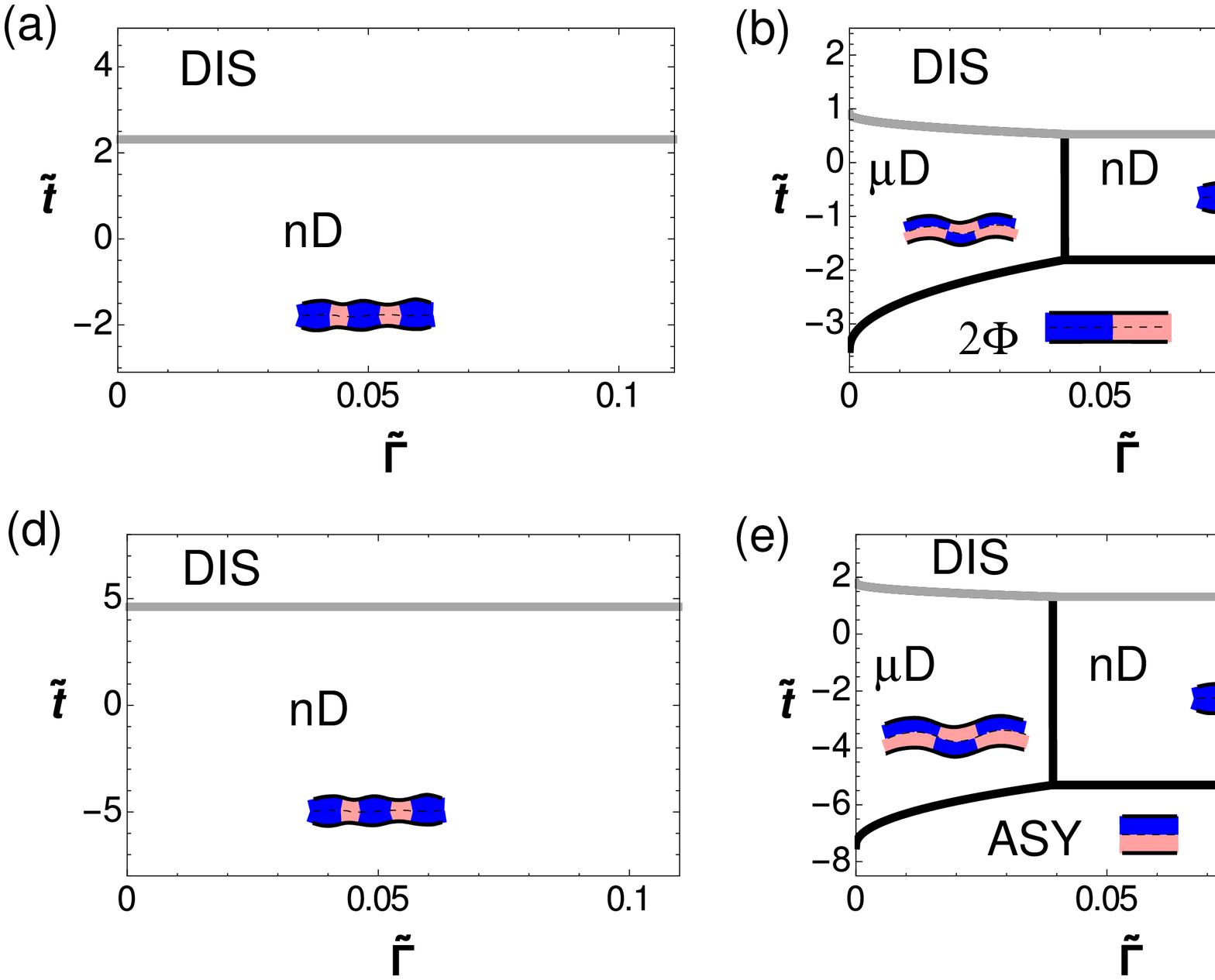}
}
\caption{
Typical phase diagrams in the plane of ''temperature'' $\tres$ and
dimensionless tension $\Gres$ for $\Lambda=0.7$ and
different choices of monolayer composition coupling $\sres$ 
curvature coupling $\Cres$.
(a) $\sres=0.1, \Cres=1.5$;
(b) $\sres=0.1, \Cres=1.$;
(c) $\sres=0.1, \Cres=0.75$;
(d) $\sres=-0.1, \Cres=2.$;
(e) $\sres=-0.1, \Cres=1.3$;
(f) $\sres=-0.1, \Cres=0.75$;
Solid gray lines correspond to continuous transitions, black lines to first
order transitions, and red arrow indicates the position of a Lifshitz point. 
}
\label{fig:phdiag_G}       
\end{figure*}

Combining all these results, we can now finally calculate the phase behavior of
the coupled membrane system. Some representative phase diagrams for membranes
with low tension are shown in Fig.\ \ref{fig:phdiag_c}. They feature the
disordered phase at high ''temperatures'' $\tres$ and a variety of ordered and
modulated states at low $\tres$. At low curvature coupling $\Cres$, the
low-temperature state is a homogeneous membrane, which is either phase
separated (2$\Phi$) for positive composition coupling $\sres$ across leaflets
(Fig.\ \ref{fig:phdiag_c}a,b,d,e), or asymmetric (ASY) for negative coupling
$\sres$ (Fig.\ \ref{fig:phdiag_c}c,f).  At high curvature coupling $\Cres$,
both the modulated microdomain phase $\mu$D and the nanodomain phase nD may
appear. Not surprisingly, the nanodomain phase tends to be favored if the
composition coupling $\sres$ between leaflets is strongly positive (Fig.\
\ref{fig:phdiag_c}a,d).  Another, less obvious factor that selects between
modulated phases is the value of the membrane parameter $\Lambda$. This can be
related to the different slopes of $\tres_{\Phi}(\Cres)$ and
$\tres_{\Psi}(\Cres)$. For small tensions, $\Gres \ll 1$, and close to the
bicritical point of $\Phi$ at $\Cres = \sqrt{\Gres}$, the slopes are given by
$\ud \tres_{\Phi}/\ud \Cres = 2 \Cres/\Gres$ and $\ud \tres_{\Psi}/\ud \Cres
\approx 2 \Cres$. Hence the nanodomain state can supersede a microdomain state
as $\Cres$ increases if $\Lambda < 1$ (Fig.\ \ref{fig:phdiag_c}a-c), the
microdomain state will remain dominant for $\Lambda > 1$ (Fig.\
\ref{fig:phdiag_c}d-f).  We recall that phospholipid bilayers have values of
$\Lambda$ around 0.7. If this finding is representative for lipid bilayers,
nanodomains would tend to be favored over microdomains at large curvature
mismatch.

Fig.\ \ref{fig:phdiag_c}b-f) also indicates regions (in the $\mu$D phase below
the dashed line), where we believe that nanodomains may exist within
microdomains, based on the adiabatic approximation described above. Such states
should be particularly interesting since they are filled with lines of
nanodomains, either regularly ordered or -- in case fluctuations destroy the
global order -- in a foam-like structure. 

Next we study the influence of tension on the phase behavior. The results for
selected parameter sets are shown in Fig.\ \ref{fig:phdiag_G}.  As a rule,
applying tension destabilizes the microdomain phase.  This is also reported in
Ref.\ \cite{SS13}. With increasing tension, the temperature range where the 
microdomain phase is stable decreases and it is eventually replaced by 
the disordered structure, a nanodomain state, or a homogeneous membrane 
structure. The phase transitions between the other states are not affected 
by membrane tension. 

We should note, however, that the tensions needed to bring about a phase
transition are rather extreme. The tension unit used to rescale the
dimensionless tension parameter $\tilde{\Gamma}$ is $\Gamma/\Gres = \sqrt{k_c
k_A}/t_0$, which corresponds to (50-100) mN/m for typical experimental values
of the elastic parameters $k_c$ and $k_A/t_0^2$ of phospholipid
bilayers\cite{Marsh06}. The tension strength where lipid bilayers rupture is
typically around 2-10 mN/m \cite{NN90}.  On short time scales, higher tensions
up to 20-30 mN/m can be sustained \cite{EHL03}. Still, the membranes will
rupture for most of the tensions considered in Fig.\ \ref{fig:phdiag_G}.  Also,
it should be noted that membranes undergo significant structural rearrangement
under high tension \cite{NWN10}, which will likely interfere with domain
formation. Hence, the phase diagrams shown in Fig.\ \ref{fig:phdiag_G} are of
rather academic interest with probably little practical relevance. The main
control that one can exert by applying tension, according to our model, is
to modulate the characteristic length scale of the microdomains. This
could again be interesting in the mixed microdomain / nanodomain state, because
it implies that tension can be used to manipulate the structural arrangements
of microdomains.

\section{Summary and discussion}
\label{sec:discussion}

In sum, we have presented a theoretical framework that allows us to describe
two curvature-driven mechanisms for the formation of modulated structures in
multicomponent lipid bilayers in a unified manner: A microdomain structure
associated with a staggered composition profile on the two monolayers, and a
nanodomain structure associated with domains that oppose each other on both
monolayers. We have seen that both microdomain and nanodomain structures can be
observed for both negative and positive couplings between the lipid
compositions on opposing monolayer leaflets. However, a strong positive
coupling favors nanodomain formation and a strong negative coupling favors
microdomain formation. We have also studied the influence of membrane tension
and found, in agreement with Ref.\ \cite{SS13}, that microdomains are
destabilized at high tensions. In our opinion, however, the most important
effect of tension is that it can be used to control the characteristic wave
length of the microdomains,

Our mean-field analysis revealed an unexpectedly complex phase behavior. It
includes two types of Bra\-zov\-skii-transitions and Ising transitions,
bicritical points where Brazovskii lines meet with Ising lines or with each
other, and possibly one Lifshitz point. We have considered the simplest
possible system, a perfectly symmetric bilayer made of identical monolayers and
a Ginzburg Landau free energy that is perfectly symmetric in the order
parameter. These are the systems most often studied in laboratory experiments.
The possible scenarios will be even more complicated if we consider asymmetric
systems where, e.g., one order parameter is favored  -- which will create a
competition between modulated phases with different symmetry \cite{KGL99} -- or
the lipid composition is different on both sides of the membranes \cite{SS13}.
Furthermore, we have assumed that only the spontaneous curvature is coupled to
the local order parameter. In reality, one would expect all elastic parameters
to depend on the local lipid composition \cite{AGF13}. In many cases, this will
probably change the picture only in a quantitative sense, phase boundaries and
length scales will be shifted. However, qualitatively new behavior may emerge
if the Gaussian modulus becomes composition dependent. In that case, the
contribution of the Gaussian curvature to the elastic free energy can no longer
be neglected, and a new type of modulated phase may emerge that might be
interesting in the context of membrane fusion.

In our mean-field approach, we have neglected fluctuations. They should have a
severe impact on the phase diagrams. It is known that thermal fluctuations turn
a continuous Brazovskii transition into a weak first order transition
\cite{Brazovskii75,HS95}, and they destabilize Lifshitz points, such that a
microemulsion channel may open up and/or part of the Ising critical line may
become first order \cite{SMS14,SMV14}.  Additionally, we have presented a
simple scaling argument in Ref.\ \cite{MVS13} according to which nanodomain
formation should preempt homogeneous demixing for {\em all} values of the
curvature coupling $\tilde{C}$. The implications of this effect for the
topology of the phase diagrams is still unclear. Most likely the Ising-type
demixing transition at low $\tilde{C}$ will persist. As mentioned in the
introduction, Ising-type demixing transitions have been observed in
multicomponent membranes and the Ising critical exponents have been verified
with great care \cite{VSK07,HCC08,HMK12}. However, the coexisting phases may be
structured and contain a certain amount of nanodomains. The Brazovskii phases
will become disordered in the vicinity of the mean-field Brazovskii line and be
replaced by a relatively strongly structured microemulsion over a wide
parameter range. 

Thus the lateral structure of the membrane is presumably characterized by a
combination of disordered micro- and nanodomains, where the nanodomains tend to
accumulate along the borders of microdomains.  Given their nanoscale size, the
nanodomains can possibly be related to the lipid ''rafts'' discussed in the
introduction. Our results indicate that microdomains can be used to manipulate
rafts, but not vice versa.  Since the size of the microdomains can be tuned by
varying the applied tension, this reveals a way how the larger lateral
organization of nanodomains or rafts could be controlled externally by a 
physical process. 

According to the raft hypothesis, raft proteins use small lipid domains as
templates, but they sometimes assemble to larger units to be functional
\cite{Pike06}. We have seen how membrane tension could be used to assist such a
process. However, our results rely on a mean-field theory and a number of
further simplifying approximations. Studying the multicomponent system in
detail by numerical simulations that also include thermal fluctuations will be
the subject of future work.

\begin{acknowledgements}

The ideas presented in this paper are based on previous work, mostly
simulations, that were carried out by Stefan Dolezel, Gerhard Jung, Olaf Lenz,
Sebastian Meinhardt, J\"org Neder, and Beate West. These simulations have given
us trust in the coupled monolayer model which on which the present model is
built.  We also wish to thank Frank Brown, Laura Toppozini, Maikel
Rheinst\"adter, and Richard Vink for collaborations that have helped to shape
our view on lipid bilayers. We thank in particular Michael Schick for helpful
comments on the manuscript and for pointing out Refs.\
\cite{GMS12,PAF13,AGF13}.

\end{acknowledgements}

\bibliographystyle{spmpsci}       
\bibliography{membrane}   

%
%

\end{document}